\documentclass[11pt]{iopart}
\usepackage{iopams}
\usepackage{epsfig} 

\def\mlambda{\mbox{\boldmath $\lambda$}}			

\def\cond{\rm cond}
\def\xhat{{\bf \hat{x}}}
\def\E{{\cal E}}
\def\EB{{\cal E}_{\rm B}}
\def\EcB{{\cal E}_{\rm c|B}}

\begin{document}

\title[]{Demodulation of RHESSI count rates by an unbiased linear Bayes estimator}
\author{Kaspar Arzner\dag
}

\address{\dag\ Paul Scherrer Institut, CH-5232 Villigen PSI, Switzerland\\
arzner@astro.phys.ethz.ch}

\begin{abstract}

The RHESSI experiment uses rotational modulation for x- and gamma ray imaging 
of solar eruptions. In order to disentangle rotational modulation from intrinsic 
time variation, an unbiased linear estimator for the spatially integrated 
photon flux is proposed. The estimator mimics a flat instrumental response 
under a gaussian prior, with achievable flatness depending on the 
counting noise. The amount of regularization is primarily given by the 
modulation-to-Poisson levels of fluctuations, and is only weakly affected
by the Bayesian prior. Monte Carlo simulations 
demonstrate that the mean relative error of the estimator 
reaches the Poisson limit, and real-data applications are shown.

\end{abstract}


\newcommand{\la}{\mbox{ \raisebox{-.5ex}{$\stackrel{\textstyle <}{\sim}$} }}
\newcommand{\ga}{\mbox{ \raisebox{-.5ex}{$\stackrel{\textstyle >}{\sim}$} }}
\newcommand{\llangle}{\langle\hspace{-0.5mm}\langle}
\newcommand{\rrangle}{\rangle\hspace{-0.5mm}\rangle}

\section{Introduction}

\begin{figure}[h]
\centerline{\epsfig{file=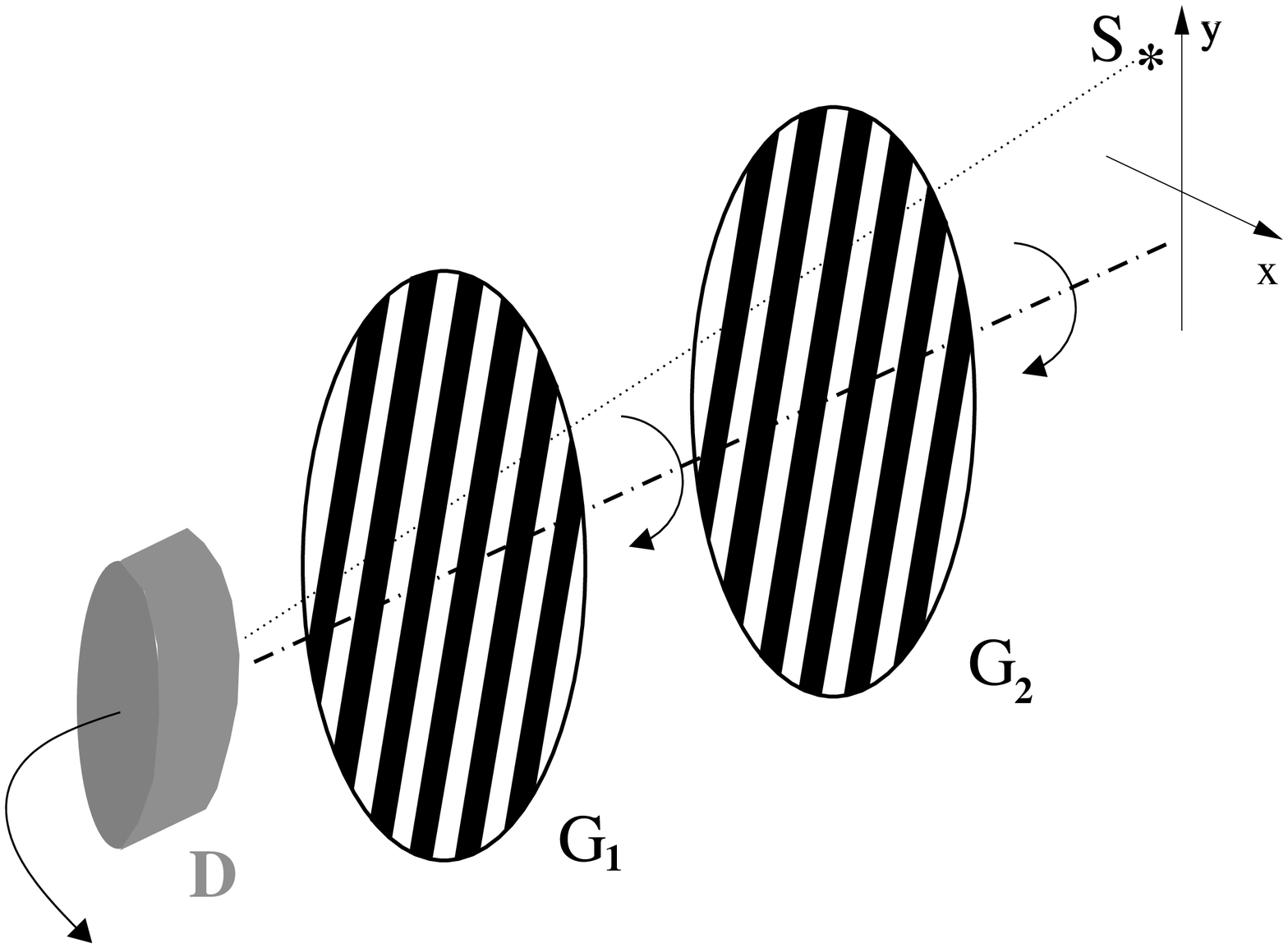,height=7cm,width=10cm}}
\vspace{-6mm}
\centerline{\epsfig{file=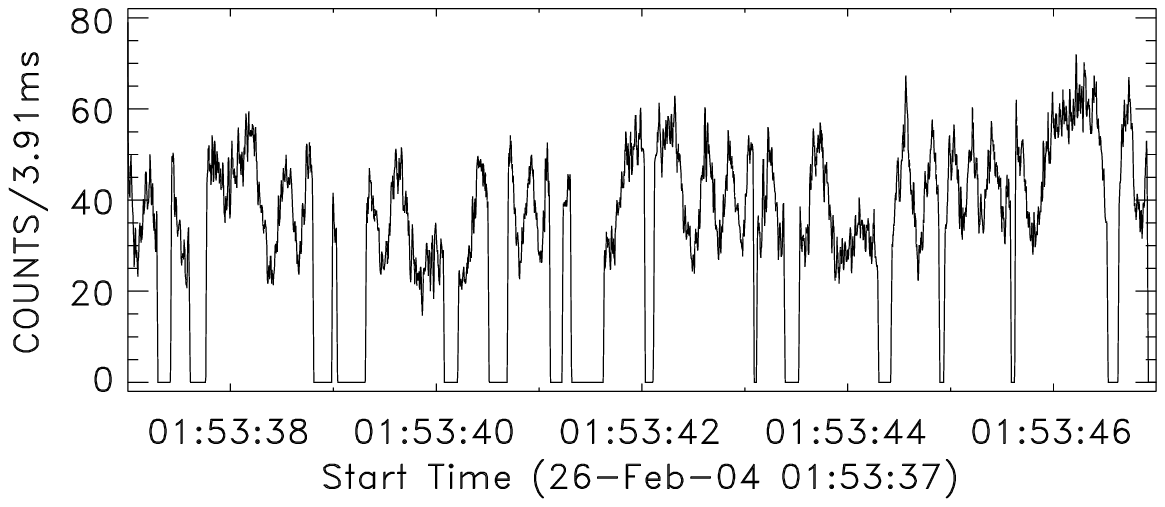}}
\caption{Rotation modulation observing principle. Top: the probability that a photon
emitted by the source (S) reaches the detector (D) is diminished if its path (dotted)
penetrates the bars (black) of the grids ($G_1$,$G_2$). As the grids rotate, the source is periodically shadowed
and released, so that the observed flux exhibits characteristic modulations (bottom), from which the true scene 
in the solar plane (x,y) can be recovered. The RHESSI instrument has 9 pairs of grids (`subcollimators'), 
of which only \#6 is shown. Times with zero count rates represent data gaps.}
\label{rmc_fig}
\end{figure}

At present, imaging of hard x-rays (HXR) and Gamma rays (GR) is only feasible by selective absorption
using different kinds of masks. An elegant and economic variant uses rotation
collimators (Schnopper 1968, Willmore 1970, Skinner and Ponman 1995) where a pair of
absorbing grids rotates between the true scene and a spatially non-resolving detector 
(Fig. \ref{rmc_fig} top). Depending on whether a source is behind or between
the grid bars, the observed flux is low or large. As rotation progresses, the true scene becomes 
thus encoded in a temporal {\it modulation} of the observed HXR/GR flux (Fig. \ref{rmc_fig} bottom).
In this process, the modulation frequency is not constant but varies with time:
glancing passages of the grid bars produce slow modulation, and rippling
passages yield fast modulation. The absolute value of the modulation frequency also
depends on the offset of the source from the rotation axis, and the modulation amplitude
depends on the source size compared to the grid period, in such a way that the amplitude is largest for 
a point source, and tends to zero if the source size exceeds the grid period.
Altogether, this results in a characteristic
time series, from which the true scene can be reconstructed by suitable inverse methods 
(Skinner 1979, Prince et al 1988, Skinner and Ponman 1995, Hurford et al 2002a). 
While such methods usually assume that the true scene does not depend on time, the present article deals 
with the complementary problem of estimating the true {\it time} dependence of the (spatially integrated) scene, 
and distinguishing it from rotation modulation. This is called here the `{\it demodulation}' problem.

The data which we envisage are obtained by the Reuven Ramaty High-Energy Solar Spectroscopic Imager 
(RHESSI, Lin et al 2002; Zehnder et al 2003). This solar-dedicated space mission uses rotation 
modulation for HXR/GR imaging of solar eruptions. The RHESSI instrument has 9 pairs of 
aligned HXR/GR-absorbing grids (9 `subcollimators') which are fixed on the rotating spacecraft.
Different subcollimators have different grid spacings, which produce sinusoidal spatial
transmission patterns (see Section \ref{rhessi_sect} for details). Behind each subcollimator there 
is a germanium detector which records the arrival time and energy (3 keV - 17 MeV)
of the incoming HXR/GR photons. From a statistical point of view,
the photon arrival times form a Poisson process with non-constant intensity. The change in
intensity is due to rotation modulation, but may also have contributions from the
time dependence of the true scene.

The latter is highly interesting from a solar physics point of view, because it is related to
particle acceleration in impulsive solar eruptions. The existence of temporal fluctuations of the 
true HXR scene down to time scales of hundred milliseconds has been confirmed by earlier 
observations (Dennis 1985, Machado et al 1993, Aschwanden et al 1993). However, these time
scales interfere with the RHESSI rotation modulation which occurs on time scales of milliseconds
to seconds. In order to disentangle time dependence of the true scene from rotation modulation
the RHESSI data must be demodulated. The most general outcome of this process would be the
true scene as a function of both space and time. But the information contained in the photon
counts is rather limited, and must be shared between spatial and temporal degrees of freedom
of the true scene. We therefore restrict ourselves here to the simpler problem of estimating the
spatially integrated true scene as a function of time. The motivation for this stems also
from the wish to compare RHESSI HXR data with spatially integrated broadband radio observations. Both
types of radiation are emitted by electrons of comparable energy, and there is some
controversy in the field whether or not the two populations actually agree. The actual
radio observations have a time resolution of 100 milliseconds, and collect radiation from 
the whole sun.

The demodulation of RHESSI data is, in general, an inverse problem. Temporal and 
spatial variations of the true scene are entangled by the observation method. An
exception only arises if the time scale of interest exceeds the RHESSI rotation period $T_S$ $\simeq$ 4s,
so that demodulation can be replaced by a running average over $T_S$. Otherwise, a good 
estimator for the spatially integrated scene requires some a priori information on its 
spatial structure in order to outweigh or damp the rotational modulation. Such information
may either come from independent observations, or from RHESSI itself. In the latter case,
one may use standard RHESSI imaging techniques (Hurford et al 2002a) to obtain an estimate of
the time-averaged true scene. These techniques assume that
the true scene is independent of time. Under this assumption, the true scene can be estimated 
after half a spacecraft rotation, when all possible grid orientations are cycled,
and higher-quality results arise from multiples of $T_S/2$. 

The aim of the present paper is to resolve time scales $\ll$ $T_S/2$.
We do not attempt here a fully general solution where
the spatially true integrated scene is found at each point of time. Instead, we will assume that the true scene
can be considered as piecewise constant during time intervals $\tau$ of order of 100 ms. Such intervals 
are shorter than most previously reported time scales, and they also agree with the time resolution
of the comparative radio observations mentioned above. The present article describes an 
efficient linear estimator for the spatially- and $\tau$-integrated scene. 
The construction of the estimator follows the lines of classical Wiener filtering (Rybicki and Press 1992),
but operates in time (not frequency) space.

The paper is organized as follows. Section \ref{rhessi_sect} summarizes some relevant technicalities of
the RHESSI instrument. Section \ref{method_sect} presents our inverse theory approach including
counting statistics (Sect. \ref{principle_sect}) and a priori information (Sect. \ref{prior_sect}). The 
resulting estimator is then discussed in Sections \ref{uniform_sect} - \ref{mechanism_sect}, 
and its performance (Sect. \ref{performance_sect}) and robustness against violation of the prior assumptions 
(Sect. \ref{robustness_sect}) are explored by Monte Carlo simulations. Section \ref{flare_sect} shows the 
algorithm at work on real data from a solar eruption. See Tab. \ref{notation} for an overview of notation.

\begin{table}[h!]
\begin{tabular}{ll}\hline
$B({\bf x},t)$	& true scene [ct asec$^{-2}$s$^{-1}$]		\\
$T_S$		& RHESSI spin period [s]			\\
$\tau$          & time interval [s] to be estimated	 	\\
$\mu = (i,j)$	& multi-index for (subcollimator, time) 	\\
$M_i({\bf x},t)$& modulation pattern				\\
$\Delta_\mu$	& time bin [s]					\\
$\lambda_\mu$	& expected counts in $\Delta_\mu$ [ct]		\\
$c_\mu$		& observed counts in $\Delta_\mu$ [ct]		\\
$b$		& expected counts in $\tau$ [ct]		\\
$\hat{b}$	& $=w_\mu c_\mu$, estimator for $b$ [ct]	\\\hline
\end{tabular}
\caption{Notation. Angular position ${\bf x}$ = ($x,y$) is measured in locally Cartesian 
heliocentric coordinates, and the dependence on ${\bf x}$ is therefore also termed `spatial'.
One arc second (1'') corresponds to 700 km on the solar disc. \label{notation}}
\end{table}

\section{\label{rhessi_sect}The RHESSI instrument}

We start with a brief description of the RHESSI response to HXRs and GRs, which defines
the `forward' problem of converting the true scene into the observed counts. We shall only 
consider photons out of a fixed energy band, average all energy-dependent quantities over
that energy band, and omit the energy dependence in the notation.

\subsection{Modulation patterns}

The instantaneous transmission probabilities of RHESSI's subcollimators, as a function of
photon incidence direction, are called the modulation patterns. They may be visualized as the 
grids' shadow on the sky plane if the detectors were operated in a transmitting mode. 
When expressed in terms of heliocentric cartesian coordinates ${\bf x}=(x,y)$ (over the limited RHESSI 
field of view, cartesian and angular coordinates are equivalent), the modulation patterns
can be approximated by

\begin{equation} 
M_i({\bf x},t) = a_{0i}(t) + a_{1i}(t) \cos \Big( {\bf k}_{i}(t) \cdot ({\bf x} - 
{\bf P}(t)) + \Psi_i(t) \Big) \, .
\label{modpat}
\end{equation}

Above, the ${\bf k}_i(t)$, $i$ = 1..9, are the grid vectors which rotate clockwise with the spacecraft 
rotation period $T_S$. All grid periods ($2\pi/|k^i| = 2.6 \cdot 3^{i/2}$ asec) are
small compared to the solar diameter (1920'') and to the fields of view of the individual subcollimators
(3600'' ... 27.000''). The coefficients ($a_{ni}(t), \Psi_i(t)$) describe the internal shadowing of the 
grids; they depend on photon energy and only weakly on time as long as the source is in
the central part of the field of view, which is the case for solar sources.
The vector ${\bf P}(t)$ is the imaging (optical) axis. It generally varies with time,
since it is not aligned with the rotation axis, which, in turn, may differ from the 
inertial axis. As a consequence, ${\bf P}(t)$ traces out a relatively complicated orbit on the solar disc,
which is continuously monitored by the onboard aspect systems (Fivian et al 2002, Hurford et al 2002b),
and the details of which need not concern us here.
All vectors ${\bf x}$, ${\bf k}(t)$ and ${\bf P}(t)$ ly in the solar plane.

\subsection{Onboard data reduction}

In order to avoid detector and telemetry 
saturation, mechanical attenuators can be inserted into the optical path, and photon counts 
can be decimated by a clocked veto: during one binary microsecond (1b$\mu$s = 2$^{-20}$s),
all events are accepted, during the next ($N_d-1$)b$\mu$s, they are all rejected. Both 
reduction methods preserve the statistical independence of the photon arriving times, 
and can be absorbed in the definition of the modulation patterns.

\subsection{\label{data_gap}Observational artifacts}

While Equation (\ref{modpat}) represents the ideal instrumental response, the real observations suffer
from several non-idealities such as detector saturation at high count rates, and sporadic breakdown
of the whole detection chain. The latter is presumably caused by cosmic rays (Smith et al 2002); 
the resulting data gaps occur at random about once per second, and have durations of milliseconds 
to seconds (see Fig. \ref{rmc_fig} bottom for an example). All non-idealities 
are combined on-ground in livetime measures 0 $\le$ $L_i(t)$ $\le$ 1, which estimate the operational 
fraction of each detector in a given time bin. The effect of livetime is taken into account by
multiplying the modulation patterns $M_i({\bf x},t)$ by $L_i(t)$. Times with $L < 0.3$ are discarded, 
which prevents detector saturation and redundant or undefined ($L=0$) numerical operations.

\section{\label{method_sect}Inverse Method}

\subsection{\label{principle_sect}Principle and treatment of measurement errors}

Let $B({\bf x},t)$ denote the solar brightness distribution at position ${\bf x}$ and time $t$,
and let the RHESSI counts be grouped in time bins $\Delta_\mu \in \tau$ where $\mu$ = ($i$, $j$) 
labels subcollimator and time. The set $\{ \mu \}$ may comprise all or only some 
of the 9 subcollimators, and different subcollimators may have different time bins. 
The observed counts $c_\mu$ in different bins $\Delta_\mu$ are supposed to be statistically independent 
Poisson variates with expectation values

\begin{equation}
\lambda_\mu = \int_{\Delta_\mu} \hspace{-2mm} dt \int d {\bf x} \, M_{i(\mu)}({\bf x},t) B({\bf x},t) \, ,
\label{lambda}
\end{equation}

where we have written $i(\mu)$ to extract the subcollimator index out of $\mu$. 
Non-solar background is neglected. The goal is to estimate spatially integrated true scene

\begin{equation}
b = \int_\tau dt \int d{\bf x} \, B({\bf x},t)
\label{b}
\end{equation}

in the time interval $\tau$, as would be observed in the absence of the subcollimators ($M_i({\bf x},t) = 1$).
The estimator $\hat{b}$ for $b$ is searched in the linear form
 
\begin{equation}
\hat{b} = w_\mu c_\mu \;\;\;\;\;\;\; \mbox{(summation convention). }
\label{hat(b)}
\end{equation}

In the sequel we shall choose $w_\mu$ to make $\hat{b}$ efficient, i.e., unbiased and minimum-variance among
all unbiased estimators. To this end, and for the sake of a concise notation, we first introduce the 
expectation operator ${\cal E}$ of a general function $X(b,{\bf c})$ of 
the true $b$ and of the observed counts ${\bf c} = \{ c_\mu \}$. This is done within a Bayesian framework where 
each true scene $B$ has assigned a prior probability $P(B)$, so that $P({\bf c},B) = P({\bf c}|B) P(B)$. One thus has

\begin{equation}
\E X(b,{\bf c}) = \underbrace{\int dP(B)}_{\EB} \, \underbrace{\sum_{{\bf c}=0}^\infty \prod_\mu \frac{e^{-\lambda_\mu} 
	\lambda_\mu^{c_\mu}}{c_\mu!}}_{\EcB} X (b,{\bf c})
\label{E}
\end{equation}

where $\lambda_\mu$ and $b$ depend on $B$ according to Equations (\ref{lambda}) and (\ref{b}).
The expectation operator (\ref{E}) involves an average over the counting statistics, followed by an
average over the prior scenes, and we write $\E = \EB \, \EcB$
to stress this sequence. If a quantity $X$ is independent of the observed counts ${\bf c}$ then 
$\E$ reduces to $\EB$. 

So far, $\EB$ is still a formal object. Let us accept 
this for the moment, and proceed with the design of the weights ${\bf w} = \{ w_\mu \}$. 
A first natural request is that the estimator (\ref{hat(b)}) is unbiased, $\E (\hat{b}) = \E(b)$, requiring that

\begin{equation}
w_\mu \, \EB (\lambda_\mu) = \EB (b) \, .
\label{unbiased}
\end{equation}

This condition locates ${\bf w}$ along the direction of $\EB(\mlambda)$. In order to find a
unique solution ${\bf w}$ we additionally require the variance ${\cal E}(\hat{b}-b)^2$ to be minimal, 
and verify a posteriori that this condition is sufficient to ensure uniqueness of the solution.
Unbiasedness and minimum-variance constraints are combined by minimizing 
$\E(\hat{b}-b)^2 + \zeta \E \, (\hat{b}-b)$ with respect to ${\bf w}$, 
where the Lagrange multiplier $\zeta$ must be adjusted to fulfill Equation (\ref{unbiased}). 
The resulting minimum condition is

\begin{equation}
\E(c_\mu c_\nu) w_\nu - \E (b c_\nu) + \zeta \E (c_\nu) = 0 \, ,
\label{min_cond1}
\end{equation}

or, after performing the average $\EcB$ over counting statistics,

\begin{equation}
\EB ( \lambda_\mu \lambda_\nu + \lambda_\mu \delta_{\mu \nu}) w_\nu - \EB (b \lambda_\nu) + \zeta \EB (\lambda_\nu) = 0 \, .
\label{min_cond2}
\end{equation}
 
Here it was used that $\EcB (c_\mu c_\nu) = \lambda_\mu \lambda_\nu + \lambda_\mu \delta_{\mu \nu}$ 
for independent counts $c_\mu$. One may now see that Equation (\ref{min_cond2})
has indeed a unique solution ${\bf w}$ if only $\lambda_\mu>0$. Under this condition, 
the matrix ${\sf \Lambda}_{\mu \nu} = \lambda_\mu \lambda_\nu + 
\lambda_\mu \delta_{\mu \nu}$ is strictly positive definite: $({\bf x}, {\sf \Lambda} {\bf x}) = ({\bf x} \cdot \mlambda)^2 + 
\sum_\mu \lambda_\mu x_\mu^2 > 0$ for all ${\bf x} \not={\bf 0}$. Since the prior
average $\EB$ represents a sum of positively weighted ${\sf \Lambda}$'s, the matrix
$\EB{\sf \Lambda}$ is also positive definite and therefore invertible -- indeed, the
condition number of $\EB {\sf \Lambda}$ can not exceed the condition number of ${\sf \Lambda}$
\footnote{see Appendix}.

Let us add at this point an interpretation of the matrix ${\sf \Lambda}$.
For any fixed realization of the true scene $B$, the fluctuations of the observed counts $c_\mu$ have two causes: 
instrumental modulation and counting noise. These two causes give rise to the two contributions $\lambda_\mu \lambda_\nu$ and 
$\lambda_\mu \delta_{\mu \nu}$ in the matrix ${\sf \Lambda}_{\mu \nu}$. At large count rates ($\lambda_\mu \gg 1)$, 
${\sf \Lambda}_{\mu \nu}$ is dominated by $\lambda_\mu \lambda_\nu$, and at small count rates
($\lambda_\mu < 1$) it is dominated by $\lambda_\mu \delta_{\mu \nu}$. The condition number of ${\sf \Lambda}_{\mu \nu}$
increases with increasing count rates and is bounded from above by $\big(|\blambda|^2 + \max(\lambda_\mu)\big) / 
\min(\lambda_\mu)$\footnotemark[1]. Thus, higher count rates
allow weaker-conditioned ${\sf \Lambda}_{\mu \nu}$. We may interpret this by saying that the 
counting noise regularizes the problem of inverting modulation, and that the amount of regularization is 
given by the level of Poisson fluctuations compared to the modulation amplitude.

\subsection{\label{prior_sect}Choice of the prior}

We turn now to the definition of $\EB$ (Eq. \ref{E}). Formally, the true brightness distribution 
$B({\bf x},t)$ is an element of a function space, and its a priori probability measure $dP(B)$ is a 
functional. In order to avoid technical complications while retaining the basic probabilistic features, 
we make the following simplifying assumptions: the true brightness distribution $B({\bf x},t)$ is concentrated
in a (known) spatial prior, where it may have (unknown) substructures which do not vary significantly 
during the time interval $\tau$ for which the unmodulated counts are to be estimated. Thus
$B({\bf x},t) \simeq B({\bf x})$ for $t \in \tau$. Now we recall that $B({\bf x})$ enters
the problem only upon weighting with the modulation paterns (Eq. \ref{lambda}). Therefore the
detailed structure of $B({\bf x})$ on scales which are small compared to the period of the modulation 
patterns do not matter, and we may represent $B({\bf x})$ by a (finite) collection of point sources

\begin{equation}
B({\bf x}) = \sum_k B_k \, \delta({\bf x} - {\bf x}_k)
\label{B_model}
\end{equation}

with the agreement that $B_k > 0$ and that the spacing between the ${\bf x}_k$ is not less
than the finest resolvable scale $l_0 \sim \min (|{\bf k}_i|^{-1})$. Since we do not wish to introduce any 
bias into the brightness distribution, except for its localization in the prior region, we set $dP(B)$ = 
$\prod_k \xi({\bf x}_k) \, d {\bf x}_k$ where $\xi({\bf x})$ is a pdf concentrated in the
prior region. This definition of $dP(B)$ is insensitive to the amplitude of $B$ but sensitive to its support.
Applying $\EB$ reduces now to an elementary calculation, 
and Equation (\ref{min_cond2}) becomes

\begin{equation}
\Big[ (1-\gamma) \langle M_\mu M_\nu \rangle + \gamma \langle M_\mu \rangle \langle M_\nu \rangle 
+ \beta^{-1} \langle M_\mu \rangle \delta_{\mu\nu} \Big] w_\nu = (
\tau - \zeta \beta^{-1}) \langle M_\mu \rangle \label{result}
\end{equation}

with 

\begin{eqnarray}
\langle M_\mu \rangle 		& = & \int_{\Delta_\mu} \hspace{-2mm} dt \, \int \hspace{-1mm} d {\bf x} 
	\, \xi ({\bf x}) M_{i(\mu)} ({\bf x},t) \label{m2} \\
\langle M_{\mu} M_{\nu} \rangle 	& = & \int_{\Delta_\mu} \hspace{-2mm} dt \, \int_{\Delta t_\nu} \hspace{-3mm} dt'
 	\int \hspace{-1mm} d {\bf x} \, \xi ({\bf x}) M_{i(\mu)}({\bf x},t) M_{i(\nu)}({\bf x},t') \label{MM2} \\
\beta 				& = & \sum_n B_n \label{beta} \\
\gamma 				& = & 1 - \beta^{-2} \sum_n B_n^2 \, . \label{gamma}
\end{eqnarray}

Equation (\ref{result}) is our main result. The parameter 0 $\le$ $\gamma$ $<$ 1 is a 
`filling factor': $\gamma$ = 0 corresponds to a single unresolved source, whereas $\gamma$ $\to$ 1 stands for 
a prior region densely covered with sources. If the diameter of the prior region exceeds the angular pitch, 
then $\gamma$ $\to$ 1 indicates the loss of modulation, while $\gamma$ = 0 retains full modulation, yet at an
unknown phase. Note that $\beta$ is the quantity which actually is to be estimated. Its occurrence 
in Equation (\ref{result}) is, however, unproblematic. On the right hand side, $\beta$ is absorbed in
the Lagrange multiplier $\zeta$. On the left hand side, $\beta$ occurs in the 
regularization only, where it may be replaced by a simpler estimate $\hat{b}_0$ (Sect. \ref{uniform_sect}) 
without qualitatively changing the solution. Since $\zeta$ only affects the scaling 
of the right hand side vector in Equation (\ref{result}), its adjustment for unbiasedness (Eq. \ref{unbiased}) 
is equivalent to re-scaling ${\bf w} \to {\bf w} \, \tau ({\bf w} \cdot \langle {\bf M} \rangle)^{-1}$,
which is easily implemented numerically.


It remains to select the pdf $\xi({\bf x})$. Our choice is largely ad hoc, and must be justified
by demonstrating that the resulting estimator $\hat{b}$ remains efficient and well-behaved even if 
the prior assumptions are violated. This will be done in Sections \ref{performance_sect} and \ref{robustness_sect}.
Here we qualitatively motivate our choice. First of all, $\xi({\bf x})$ should be simple and 
depend on not more than a few parameters in order to facilitate operational data processing.
One set of parameters which is provided by the RHESSI data products is the centroid ${\bf x}_0$ of 
the HXR emission derived from time-integrated ($\gg \tau$) RHESSI imaging.
It is natural to center $\xi({\bf x})$ on ${\bf x}_0$. In addition to the centroid we would like to specify 
the rough size $l$ of the prior region, but not any of its details, since these are not known on time scales
as small as $\tau$. The choice of $l$ reflects the uncertainty of the true scene, and
may be based on RHESSI imaging or on independent observations. In particular, we may chose $l$ such as to 
cover a solar `active region' (Howard 1996) derived from magnetic field observations. In any case,
$l$ should not be unrealistically small; simulations 
(Sect. \ref{robustness_sect}) suggest that $l > 10''$ is a reasonable, conservative, choice.
Owing to the inherent rotation symmetry of the rotation modulation observing principle, it is 
suggestive to choose an isotropic form for $\xi({\bf x})$, although this is by no means a rigorous
request. From a practical point of view it is also important that the integrals 
in Equation (\ref{m2}) and (\ref{MM2}) can be performed analytically as far as possible
to save numerical operations. A choice which fulfills the above criteria is

\begin{equation}
\xi({\bf x)} = (2 \pi l^2)^{-1} e^{-({\bf x}-{\bf x}_0)^2/2l^2} \, .
\label{prior}
\end{equation}

Since the coefficients $a_{ni}(t)$ and $\Psi_i(t)$ in Equation (\ref{modpat}) are slowly
varying with time, they may be replaced by discrete-time versions $a_{n\nu}$ and $\Psi_\nu$.
Equations (\ref{m2}, \ref{MM2}) then become

\begin{eqnarray}
\langle M_\mu \rangle & = & \sum_{m=0}^{1} a_{m\mu} e^{-\frac{l^2}{2} |m {\bf k}_{i(\mu)}|^2} 
		\int_{\Delta_\mu} \hspace{-2mm} dt  \cos \big( m \phi_\mu \hspace{-.5mm} (t)\big) \label{mc} \\[2mm]
\langle M_\mu M_\nu \rangle & = & \sum_{s=\pm1} \sum_{m,n=0}^{1} \frac{a_{m\mu} a_{n\nu}}{2} 
			e^{-\frac{l^2}{2}(|m{\bf k}_{i(\mu)}|^2+|n{\bf k}_{i(\nu)}|^2)}\int_{\Delta_\mu} 
			\hspace{-2mm} dt \int_{\Delta_\nu} \hspace{-2mm} dt' \times \label{mmc} \\
		&  & \times  e^{-l^2 smn {\bf k}_{i(\mu)}\hspace{-.5mm}(t) \cdot {\bf k}_{i(\nu)}\hspace{-.5mm}(t')} 
		\cos \big( m\phi_\mu \hspace{-.5mm}(t) + sn\phi_\nu \hspace{-.5mm}(t') \big) \nonumber
\end{eqnarray}

with $\phi_\mu$$(t)$ $ = $ ${\bf k}_{i(\mu)}(t) \cdot ({\bf x}_0-{\bf P}(t)) + \Psi_\mu$. On time scales $\tau \ll T_S/2$, 
mostly terms with $s$=$-1$ contribute to Equation (\ref{mmc}). The time integrals in
Equations (\ref{mc}) and (\ref{mmc}), which depend on the actual spacecraft motion,
are evaluated numerically.

\subsection{\label{bin_sect}Choice of time bins}

RHESSI detects individual photons, and their assignment to time bins
is an important first step of the demodulation procedure. There are two, potentially conflicting,
requests on the time bins $\Delta_\mu$. On the one hand, $\Delta_\mu$ should resolve (say, by a factor 10)
the instantaneous modulation period
\begin{equation}
\tau_{i}^{\rm mod}(t) \sim T_S \, |{\bf k}_{i}(t) \times \big({\bf x}_0-{\bf P}(t) \big)|^{-1} \;\;\;\; [s]
\label{tau_mod}
\end{equation}
of a source at position ${\bf x}_0$.
On the other hand, demodulation is only beneficial if there are sufficient counts to observe modulation
against Poisson noise. This requires, empirically,
some 5 counts per time bin. In addition, the time bins should be integer 
fractions of $\tau$ in order to minimize roundoff error. We therefore adopt the following rule:

\begin{equation}
\Delta_\mu = \frac{\tau}{1 + {\rm round} \, (\Delta_\mu^*/\tau)} \; , \;\;\;\;\; 
\Delta_\mu^* = \max \, \bigg( \frac{5}{\langle c_{i(\mu)} \rangle}, \,  \frac{1}{10} \min_{t \in \tau} \tau^{\rm mod}_{i(\mu)}(t)  \bigg)
\label{choice_of_Delta_t}
\end{equation}

where $\langle c_i \rangle$ is the average count rate [ct/s] in subcollimator $i$. The factors 5 and 10 in Equation
(\ref{choice_of_Delta_t}) are empirical.

\subsection{\label{num_sect}Numerical implementation}

Equations (\ref{mc}) and (\ref{mmc}) are evaluated by an extended trapezoidal rule
with intermediate step size adapted to the modulation frequency and amplitude. The inversion of the
matrix on the left hand side of Equation (\ref{result}) is performed by singular value 
decomposition (Golub and Van Loan 1989) with explicit control of condition number.

\section{\label{discussion_sect}Discussion}

\subsection{\label{uniform_sect}Limiting behaviour}

Let us start our discussion by verifying that the estimator (\ref{hat(b)}) remains well-behaved and meaningful 
in extreme cases where the prior is pointlike or flat, and where the observed flux tends to zero. The limit
of infinite flux is discussed in Sect. \ref{geom_sect}.

At very low count rates ($\ll$ 1 ct/$\tau_i^{\rm mod}$), modulation is no longer observable, and
one may thus expect that $\hat{b}$ should reduce to a simple average, which is an
efficient estimator for the pure counting noise problem. Indeed, in the limit $\beta$ $\to$ 0, 
Equation (\ref{result}) yields (after adjusting $\zeta$ to satisfy Eq. \ref{unbiased})

\begin{equation}
\hat{b} \to \hat{b}_0 \doteq \frac{\tau \sum_\mu c_\mu}{\sum_\mu \langle M_\mu \rangle}
	\simeq \frac{\tau \sum_\mu c_\mu }{\sum_\mu \Delta_\mu a_{0\mu}} \, .
\label{uniform}
\end{equation}


Equation (\ref{uniform}) will be referred to as `uniform average' since it involves uniform weights ($w_\mu$=const).
The last approximation in Equation (\ref{uniform}) holds if the modulation
is fast ($\tau^{\rm mod}_i \ll \tau$) or weak ($|{\bf k}_i| l \gg 1$), so that the term with $m$=1
in Equation (\ref{mc}) can be neglected. The uniform average $\hat{b}_0$ provides 
a simple guess of $b$, which is - hopefully -- improved by the more sophisticated
estimator (\ref{hat(b)}). We shall see in Sect. \ref{performance_sect} - \ref{robustness_sect} that 
this is in fact the case.

The uniform average (\ref{uniform}) is also attained if $l$ = 0 or $\gamma$ = 1, both representing
deterministic limits with completely localized or densely filled prior regions.
In either case, the term $(1-\gamma)\langle M_\mu M_\nu \rangle + \gamma \langle M_\mu \rangle \langle M_\nu \rangle$
reduces to $\langle M_\mu \rangle \langle M_\nu \rangle$, so that Equation 
(\ref{result}) has the solution $w_\mu$ = const, and $\hat{b}$ reduces to $\hat{b}_0$.

Another situation of interest is $l \to \infty$ where the prior becomes globally flat. For $l \to \infty$ one can derive from Equations 
(\ref{mc}-\ref{mmc}) that $\langle M_\mu \rangle \to a_{0\mu} \Delta_\mu$ and $\langle M_\mu M_\nu \rangle 
\to a_{0\mu} \Delta_\mu a_{0\nu} \Delta_\nu + \frac{1}{4} (a_{1\mu} \Delta_\mu)^2 \delta_{\mu \nu}$. 
Thus Equation (\ref{result}) can be inverted by the Sherman-Morrison formula (Press et al 1998), and one finds 
(after adjusting $\zeta$)

\begin{equation}
\hat{b} \to \hat{b}_{\rm fl} \doteq \frac{\tau \sum_\mu c_\mu \sigma_\mu^{-2} }{\sum_\mu \Delta_\mu a_{0\mu} \, \sigma_\mu^{-2}}
\;\;\;\; \mbox{where} \;\;\; \sigma_\mu^2 = 1 + \frac{1}{4} (1-\gamma) \, \beta \Delta_\mu \, \frac{a_{1\mu}^2}{a_{0\mu}} \, .
\label{flat}
\end{equation}

From Equation (\ref{flat}) the uniform average (\ref{uniform}) is recovered
if $\frac{1}{4} (1-\gamma) \beta \Delta_\mu \frac{a_{1\mu}^2}{a_{0\mu}} \ll 1$. Otherwise a correction arises
which weakly varies with time (Sect. \ref{rhessi_sect}). Bins with large modulation amplitude $a_{1\mu}$ are 
given less weight in Equation (\ref{flat}), which is reasonable since their uncertainty is larger.

In summary, we have found that the estimator (\ref{hat(b)}) has well-defined and meaningful analytic
limits when the flux tends to zero ($\beta \to 0$), and when the prior is completely restrictive ($l \to 0$) or 
completely nonrestrictive ($l \to \infty$).

\subsection{\label{geom_sect}Geometrical interpretation}

In the limit of infinite count rates, the counting noise becomes unimportant and Equation (\ref{result}) admits a purely
geometrical interpretation. Recalling the assumption $B({\bf x},t) \simeq B({\bf x})$
for $t \in \tau$, one has that $\EcB (b) = \int d {\bf x} \, B({\bf x}) \, \Phi({\bf x})$ with 
$\Phi(x) \doteq w_\mu M_\mu({\bf x})$ and $M_\mu({\bf x}) \doteq \int_{\Delta_\mu} \hspace{-1mm} dt \, M_{i(\mu)}({\bf x},t)$.  
Therefore $\EcB (b)$ is a good estimator for $b$ for {\it arbitrary} 
$B({\bf x})$ if $\Phi({\bf x})$ is independent of ${\bf x}$. The modulation patterns would
then be canceled. However, the functions $M_\mu({\bf x})$ are neither orthogonal nor complete, so that $\Phi({\bf x})$ cannot be 
made constant by whatever choice of weights $w_\mu$. Instead, we may try
to minimize the fluctuations of $\Phi({\bf x})$ within the spatial prior $\xi({\bf x})$, and therefore
minimize $\int \xi({\bf x}) (\Phi({\bf x})-{\rm const})^2 \, d {\bf x}$. This yields 
$\langle M_\mu M_\nu \rangle w_\nu \propto \langle M_\mu \rangle$. Comparing this expression
to Equation (\ref{result}) we see that Equation (\ref{result}) minimizes
the fluctuations of $\Phi({\bf x})$ within the spatial prior under the maximum-modulating 
assumption that there is a single point source ($\gamma=1$) observed at infinite count rate 
($\beta \to \infty$). 

The matrix ${\sf M} = \langle M_\mu M_\nu \rangle$ is generally ill-conditioned,
and the weights ${\bf w} = {\sf M}^{-1} \langle {\bf M} \rangle$ therefore oscillate.  
At finite count rate, the excursions to large $|w_\mu|$ amplify the Poisson noise, and
thereby impair the estimator $\hat{b}$ (Eq. \ref{hat(b)}). An optimum estimator must thus 
limit the oscillations of ${\bf w}$ to a level which commensurates with the Poisson noise. 
Equation (\ref{result}) provides a possible trade-off.

\begin{figure}[h]
\hspace{-15mm}
\epsfig{file=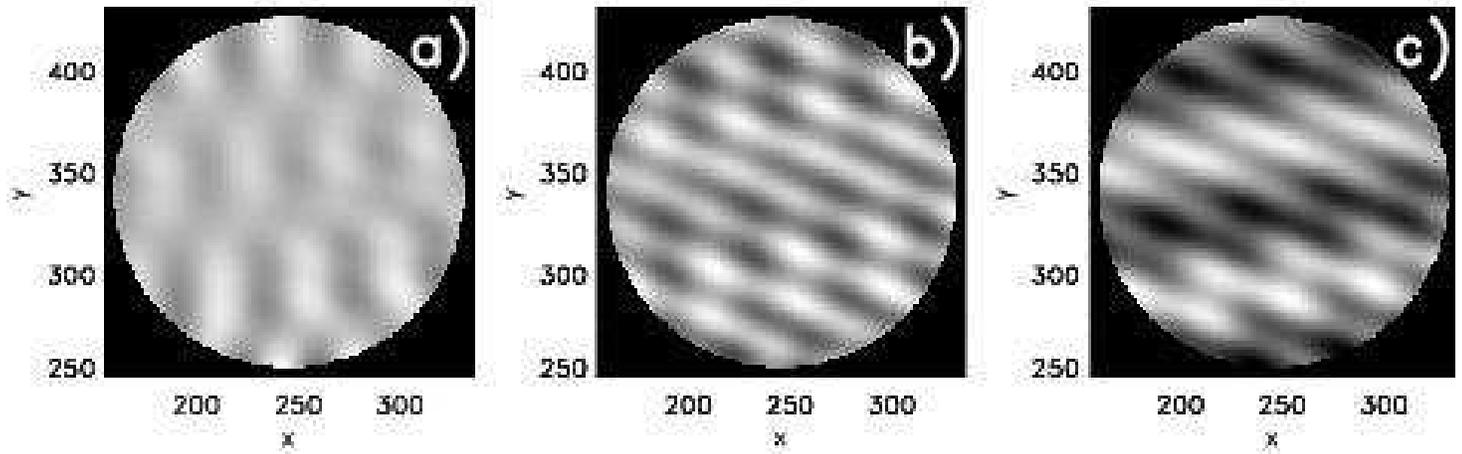}
\caption{The function $\Phi({\bf x})$ obtained from subcollimators (3,4,5,6) and time interval $\tau$=0.2s
starting at Feb. 26 2004, 01:53:39.850 UT. a) for infinite count 
rate ($\Delta \Phi/\Phi$=0.0058); b) adapted to the actual count rate ($\Delta \Phi/\Phi$=0.024); 
c) in the limit of zero count rate ($\Delta \Phi/\Phi$=0.14). For 
$\Delta \Phi / \Phi \to 0$ the effect of modulation is canceled. See text.}
\label{phi_fig}
\end{figure}

Fig. \ref{phi_fig} illustrates the resulting function $\Phi({\bf x})$ for $\gamma$ = 0 and different $\beta$, using
aspect data of February 26 2004, subcollimators (3,4,5,6), and $\tau$ = 0.2s. 
The time bins $\Delta_\mu$ are (2.8, 4.8, 8.5, 15)ms, and
the spatial prior (size $l$ = 40'') is centered at the source position (245'', 340'') estimated
from time-integrated RHESSI imaging. The black mask in Fig. \ref{phi_fig} indicates the 10\% level of 
the spatial prior (Eq. \ref{prior}). In order to assess the constancy of $\Phi({\bf x})$ we consider 
the standard deviation $\Delta \Phi$ and mean $\Phi$ inside the black mask.
In a noiseless world ($\beta \to \infty$), the modulation patterns would admit $\Delta \Phi / \langle \Phi \rangle \sim$ 
0.0058 (Fig. \ref{phi_fig}a). Taking the actual counting noise ($\beta \sim \hat{b}_0 = $ 6200 ct/s) into 
account, $\Delta \Phi / \langle \Phi \rangle \sim$ 0.024 is still
achievable (Fig. \ref{phi_fig}b). For comparison, the uniformly weighted ($w_\mu$=const) case is
also shown (Fig. \ref{phi_fig}b, $\Delta \Phi / \langle \Phi \rangle \sim$ 0.14), which
corresponds to the limit $\beta \to 0$. As can be seen, the function $\Phi({\bf x})$ becomes less
efficient in canceling the modulation patterns when the count rate $\beta$ decreases. But, 
at the same time, the sensitivity of $\hat{b}$ to Poisson noise increases. The trade-off chosen by
the present method is shown in Fig. \ref{phi_fig}b.

\subsection{\label{mechanism_sect}Weighting mechanism}

In a generic situation ($\gamma$$\not=$1, $l$$\not$=0, $\hat{b}_0$$\gg$1) one may identify three mechanisms 
by which the weights operate. These are most easily discussed in terms of Fourier modes of the modulation.
First, if the modulation phase is resolved ($|{\bf k}_i| l \ll 1$) then modulation can `actively' be 
countersteered. Secondly, if the modulation phase is not resolved but the instantaneous modulation frequency 
$1/\tau_i^{\rm mod}$ (Eq. \ref{tau_mod}) is known, then counts with $\tau_i^{\rm mod} \not\ll \tau$ 
can be suppressed because they do not allow a `passive' averaging. This suppression works
for each subcollimator individually. (Although $\tau_i^{\rm mod}$ does not explicitly 
show up in Equation (\ref{result}), it is implicitly contained in the diagonal blocks of
correlation function $\langle M_\mu M_\nu \rangle$, see Fig. \ref{mm_fig} below). Thirdly,
the matrix $\langle M_\mu M_\nu \rangle$ couples different subcollimators, which regulates
the relative weighting of different subcollimators.

\begin{figure}[h]
\centerline{\epsfig{file=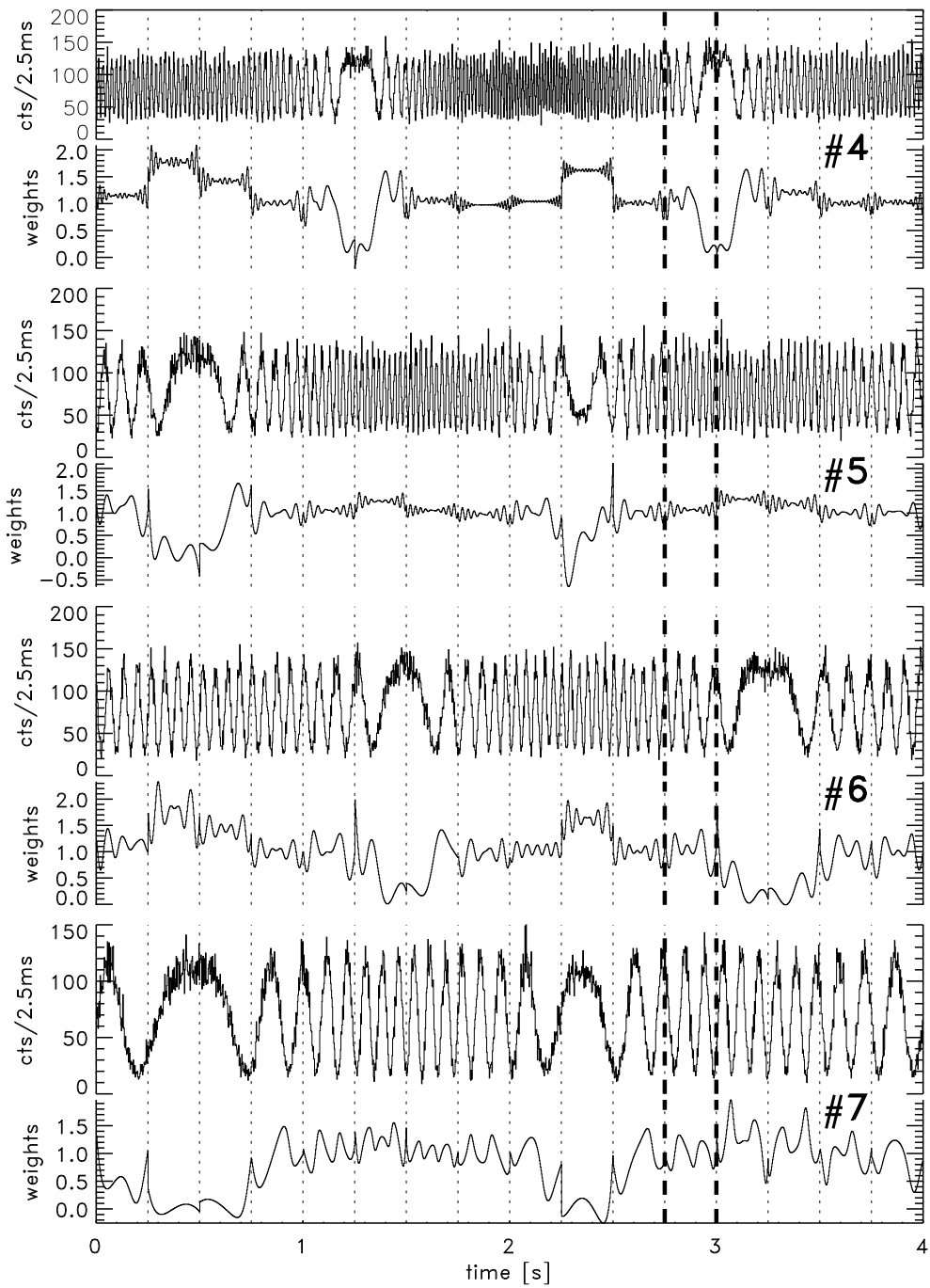}}
\caption{Simulated example of binned count rates and corresponding weights in
disjoint intervals (dotted). The boldface dashed interval 
refers to Fig. \protect\ref{mm_fig}. See text.}
\label{cw_fig}
\end{figure}

The different mechanisms are illustrated in Figures \ref{cw_fig} and \ref{mm_fig}, showing simulated 
data of subcollimators (4,5,6,7) with ${\bf x}_0$ = (885'',161''), $l$ = 20'', $\gamma$ = 0.2, 
and a single intense ($\sim$ 3$\times$$10^4$ ct/s/subcollimator) point source located at ${\bf x}_0$.
The chosen geometry represents a solar limb source. The simulated counts are divided into 
disjoint intervals of duration $\tau$ = 0.25s (Fig. \ref{cw_fig}), and time bins are 
equal ($\Delta_\mu$ = 0.0025s) for better comparability. The outweighing of modulation 
is most clearly seen in the coarsest subcollimator (Fig. \ref{cw_fig}, \#7) with
$l/p_7$ = 0.16, where $p_7 = 2 \pi / |{\bf k}_7|$ = 122'' is the period of subcollimator \#7.
The modulation is also -- though less efficiently -- outweighed in subcollimator \#6 ($l/p_6$ = 0.28), 
while the finest subcollimators \#5 ($l/p_5$ = 0.49) and \#4 ($l/p_4$ = 0.85) mostly
operate in an averaging mode, with priods of low $\tau^i_{\rm mod}$ being suppressed.

The different regimes manifest in different forms of the correlation $\langle M_\mu M_\nu \rangle$ 
(Fig. \ref{mm_fig}). For $l/p_i$ $\ll$ 1 the matrix 
$\langle M_\mu M_\nu \rangle$ approximately factorizes into $\langle M_\mu \rangle\langle M_\nu \rangle$ 
(Fig. \ref{mm_fig}, \#7). With increasing $l/p_i$, the autocorrelation of
a single subcollimator takes the form $\langle M_{(i,t)} M_{(i,t')} \rangle
\sim \cos (2 \pi \frac{t-t'}{\tau_i^{\rm mod}})e^{-(t-t')^2/2(\tau_i^{\rm d})^2}$ with decay 
time $\tau_i^{\rm d}$ $\sim$ $(lk_i)^{-1}({\bf \dot{k}}_i \cdot {\bf x}_0)^{-1}$ (Eq. \ref{mmc}).
Here, both $\tau_i^{\rm d}$ and $\tau_i^{\rm mod}$
depend to first order on $(t+t')/2$, which results in the `chirping' behaviour of
Fig. \ref{mm_fig}, subcollimator \#4. The chirp towards low modulation frequencies
is associated with a drop of the corresponding weights (Fig. \ref{cw_fig} top, time
interval in dashed lines).

\subsection{\label{performance_sect}Performance}

\begin{figure}
\centerline{\epsfig{file=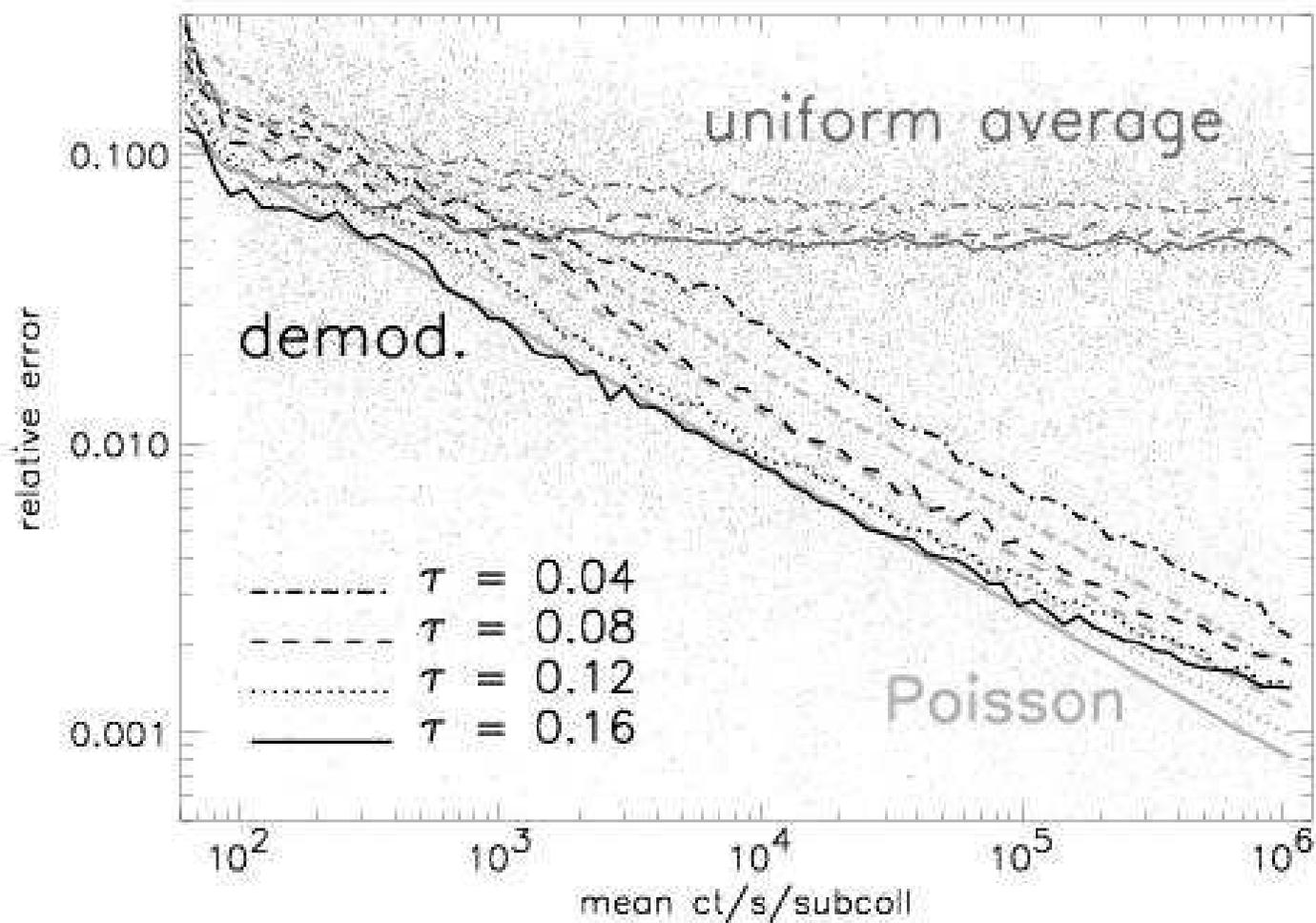}}
\caption{Simulated relative errors of demodulation, uniform average, and the Poisson limit. 
The curves represent averages over $10^5$ samples, some of which are shown as dots. All subcollimators 
are used, $\gamma$ = 0.1, and $l$ = 40''. A color rendering of this figure is available in the 
online version.}
\label{performance_fig}
\end{figure}

The quality of the demodulation and its robustness against violation of prior assumptions have
been explored by Monte Carlo simulations. Figure \ref{performance_fig} shows, as an example, the relative 
errors $|b-\hat{b}|/b$ of $10^5$ simulated time intervals $\tau$, together with uniform values $|b-\hat{b}_0|/b$ 
and the relative Poisson error $N_{\rm tot}^{-1/2}$ with $N_{\rm tot}$ the total number of counts
in $\tau$. Dots illustrate a subsample of the simulation; graphs represent the full ensemble
averages. The spatial prior is centered at ${\bf x}_0$ = (420'', -630'') and has size 
$l$ = 40''. The simulated brightness distribution is a superposition of 10 random sources
$A_s \exp \{ -\frac{1}{2} (t-t_s)^2/\tau_s^2 -\frac{1}{2} ({\bf x}-{\bf x}_s)^2/\l_s^2 \}$
with uniform relative amplitudes $A_s/\sum_sA_s$, Gaussian positions ${\bf x}_s$ (mean ${\bf x}_0$, variance $l^2/4$),
uniform sizes $l_s$ $\in$ (1''$\,...\,$5''), and uniform intrinsic time scales $\tau_s$ $\in$ (0.2s$\,...\,$1s).
The assumed filling factor is $\gamma$ = 0.2, whereas the true value scatters from 0.8 to 0.9. Data gaps are 
neglected. At low count rates, the error is dominated by the Poisson noise, and all three types of error coincide. 
At higher count rates, the error becomes dominated by modulation, and the uniform average no 
longer improves with improving counting statistics. The estimator $\hat{b}$
performs better, and reaches the Poisson limit for integration times $\tau$ $\ge$ 
8ms and count rates up to $10^5$ ct/s/subcollimator. The latter is, in fact,
the practically relevant range -- higher count rates do not occur because the RHESSI detectors 
need $\la$ 8$\times$$10^{-6}$s to recover after each photon impact. One may thus be confident 
that -- at  least with regard to the (tolerant) error measure ${\cal E} ( |b-\hat{b}|/b )$ --
the estimator $\hat{b}$ works optimally for practical purposes.
The degradation of $|b-\hat{b}|/b$ at highest count rates and largest $\tau$ is attributed 
to the increased susceptibility to prior assumptions, and possibly also to numerical errors as the 
involved matrices become large and weak-conditioned. (Condition numbers up to $10^5$ occur at
$10^6$ ct/s/subcollimator.) Extended simulations including different brightness distributions, 
data gaps, and $\gamma$ values yielded similar results.

\subsection{\label{robustness_sect}Influence of the prior}

\begin{figure}
\centerline{\epsfig{file=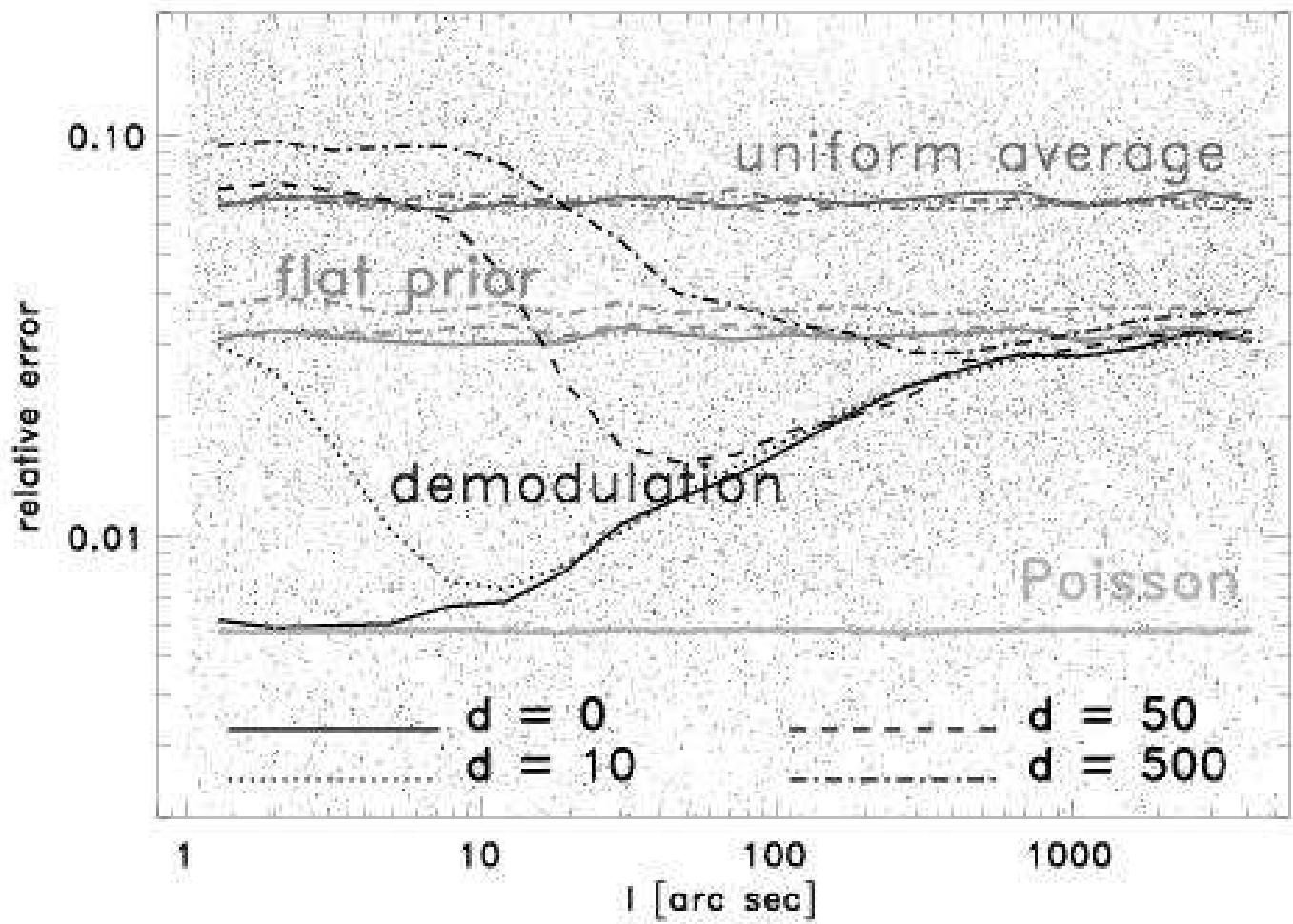}}
\caption{Influence of the prior on the demodulation.
The true source is $d$ arc seconds away from the prior centroid. The simulation explores
different prior sizes $l$ at fixed count rate $2 \cdot 10^4$ ct/s/subcollimator and time
interval $\tau$ = 0.12s, and displays relative errors similar as in Fig. \protect\ref{performance_fig}. 
The `demodulation' is given by Eq. (\protect\ref{result}); the `uniform average' by Eq. (\ref{uniform}); the
`flat prior' by Eq. (\ref{flat}). A color rendering of this figure is available in the online version.}
\label{robustness_fig}
\end{figure}

An important practical feature of Bayesian inverse methods is their robustness against violation of the 
prior assumptions. If a wrong prior is used, then the result will be degraded but it should
not become (much) worse than if no prior was used at all (i.e., if the weights were uniform or
the prior was flat, $l \to \infty$).
We shall now demonstrate that this goal is met by our demodulation method. Figure 
\ref{robustness_fig} shows simulations performed at fixed count rate of $2 \cdot 10^4$ ct/s/subcollimator
and fixed integration time $\tau$ = 0.12s, but with varying size $l$ of the prior region, and with different offsets
between the true and prior centroids. In each simulation, a prior centroid ${\bf x}_0$ is 
chosen at random across the solar disc, and the true centroid is placed $d$ arc seconds away in
random direction. The true source has several gaussian components,
which are all concentrated in a narrow (3'') region around the true centroid. The simulated imaging 
axis ${\bf P}(t)$ moves within the central ($<$200'') region of the solar disc.
While the cases $d=(0,10,50)''$ reflect realistic uncertainties of RHESSI observations, 50'' being a 
conservative value, the case $d=500''$ is overly pessimistic 
and is included here for demonstration purposes. The relative errors (vertical axis) are defined as in
Figure \ref{performance_fig}; the `flat prior' estimator is given by Equation (\ref{flat}).
Dots again represent a subsample of the simulation, while curves represent 
averages over the full sample ($6 \cdot 10^4$).

If the prior centroid coincides with the true one ($d$=0'', black solid line), then the demodulation
reaches the Poisson limit for $l \la$ 5'', i.e., as long as the prior width does not exceed the true
width by more than a factor $\la$2. If the true and assumed centroids differ ($d>0$), then the demodulation
does no longer reach the Poisson limit, and degradation depends on the ratio $d/l$. As can be seen, 
all error curves collapse to the optimum ($d=0$) one when $l \ga d$, indicating that discrepancies
between true and prior centroids are tolerated up to the size of the prior region.
Except for very large and unrecognized prior errors ($d$=500'', $l$ $<$ 20''),
the demodulation performs better than the uniform average, and approaches the flat prior estimate 
as $l \to \infty$. The flat prior, in turn, performs better than the uniform average.
The solar diameter is 1920'', so that $l > 2000$'' effectively represents a trivial prior, which only 
indicates that the photons came from the sun.

As a practical implication, we learn from Figure \ref{robustness_fig} (and from similar simulations) 
that, for realistic uncertainties (say, $d \le 50''$), demodulation is beneficial compared to 
the uniform average if $l \ga 10''$, and it is beneficial compared to the flat prior if $l \ge 20''$.

\subsection{\label{flare_sect}Example of a solar eruption}

\begin{figure}
\centerline{\epsfig{file=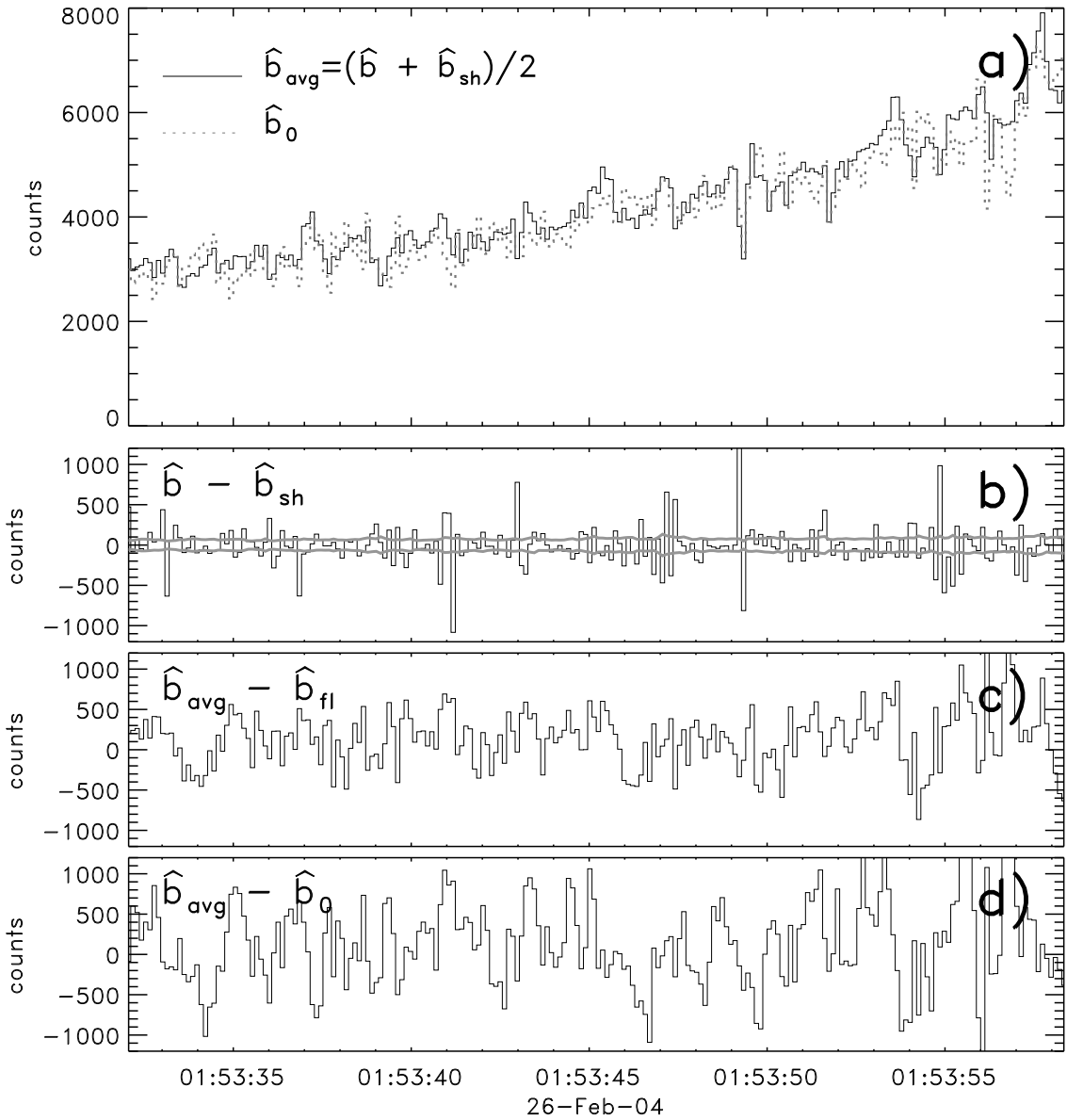}}
\caption{Real-data application of demodulation. a) demodulation (solid) and uniform average (dotted) 
of subcollimators (1,3,4,5,6,7,8,9) and $\tau$ = 0.12s. The demodulation $\hat{b}_{\rm avg}$ is an average 
over two solutions $\hat{b}$ and $\hat{b}_{\rm sh}$ (Eqns. \protect\ref{hat(b)}, \protect\ref{result})
which differ only by an offset $\tau/2$ of the time intervals. b) discrepancy of the two 
solutions, together with the pure Poisson error (gray line). c) residuals between
$\hat{b}_{\rm avg}$ and the flat prior estimate $\hat{b}_{\rm fl}$ (Eq. \protect\ref{flat}).
d) residuals between $\hat{b}_{\rm avg}$ and the uniform average $\hat{b}_0$ 
(Eq. \protect\ref{uniform}).}
\label{flare_fig}
\end{figure}

A recent solar eruption, to which also Figures \ref{rmc_fig}--\ref{phi_fig} refer, occurred on February 
26, 2004. RHESSI has observed the whole eruption, and Figure \ref{flare_fig} shows a part of the impulsive 
rise phase, where most temporal fine structures are expected. The observed counts of subcollimators (1,3,4,5,6,7,8,9)
are divided into disjoint intervals of duration $\tau$ = 0.12s, while subcollimator 2 is rejected 
due to increased background (Smith et al. 2002). The centroid ${\bf x}_0$ =
$(245'',340'')$ and size $l$ = 30'' of the spatial prior are taken from a long-exposure (10s) 
RHESSI imaging and in agreement with the simulations of Sect. \ref{robustness_sect}.
The filling factor is assumed to be $\gamma$ = 0.4. The average count rate is about
4000 ct/s/subcollimator, and energies from 5 to 15 keV are used.

In order to remove some arbitrariness of the time binning and to 
assess the quality of the demodulation $\hat{b}$ we proceed as follows. By assumption,
the true scene is approximated as piecewise constant in time. If this assumption was true then a second
demodulation $\hat{b}_{\rm sh}$ with intervals of equal duration $\tau$ but shifted by $\tau/2$ 
should yield a similar result. By comparing $\hat{b}$ with $\hat{b}_{\rm sh}$ we may thus gain an 
estimate on the accuracy of the demodulation, and by considering $\hat{b}_{\rm avg} = \frac{1}{2}(\hat{b}
+\hat{b}_{\rm sh})$ we may remove some arbitrariness of the time binning.

Figure \ref{flare_fig} (top panel) shows the estimator $\hat{b}_{\rm avg}$, together with the
uniform average $\hat{b}_0$ as the simplest possible guess. Panel b) shows the
discrepancy between $\hat{b}$ and its time-shifted version $\hat{b}_{\rm sh}$. As can be seen,
the relative discrepancy $|\hat{b}-\hat{b}_{\rm sh}|/\hat{b}_{\rm avg}$ is throughout small (6\%). The pure Poisson
error $\sqrt{\sum_\mu c_\mu w_\mu^2}$ is also shown for comparison (gray line). The residuals
$\hat{b}-\hat{b}_{\rm sh}$ exceed the pure Poisson error, and this excess is due
to uncertainties of the weights ${\bf w}$. The latter are caused by the principal reasons
discussed in Sect. \ref{geom_sect}, but also have contributions from the uncertainty of
the prior centroid ${\bf x}_0$ and -size $l$, as well as from the approximate form of Equation 
(\ref{modpat}) and its energy-averaged coefficients $\{ a_{ni}(t), \psi_i(t) \}$, possible
errors of the aspect data $\{ {\bf P}(t), \arg {\bf k}_i(t)\}$,
and from violation of the piecewise constant-in-time approximation of
the true scene. A complete disentangling of different sources of errors is difficult and
somewhat speculative. However, we may empirically compare the residuals $\hat{b} - \hat{b}_{\rm sh}$
to the residuals $\hat{b}_{\rm avg} - \hat{b}_{\rm fl}$ (Fig. \ref{flare_fig}c) and $\hat{b}_{\rm avg} - \hat{b}_0$
(Fig. \ref{flare_fig}d). This shows a clear order of residual amplitudes b) $<$ c) $<$ d),
in agreement with the simulations. We may thus be confident that the flat prior estimate improves
the uniform average, and that the demodulation improves the flat prior estimate.
All residuals are centered about zero, in agreement with unbiasedness. By comparing simulation results
with the residuals of Fig. \ref{flare_fig} b-d), and with families of similar real-data demodulations 
with varying $\tau$ (not shown), we conclude that the demodulation error
is in the order of the residuals $\hat{b}-\hat{b}_{\rm sh}$ (Fig. \ref{flare_fig}b).
As may be seen from Figure \ref{flare_fig}a), many of the
excursions of $\hat{b}_0$ -- especially towards low count rates (data gaps) --
are absent in $\hat{b}_{\rm avg}$. They are therefore most likely to be instrumental.
Not all of the data gaps can, however, be removed by the demodulation: see, e.g., before 01:53:50. Here,
the collective dropout of several detectors during $> \tau$ inhibits successful compensation.

\section{Summary}

We have developed a unbiased linear Bayes estimator for the photons arriving in front of the RHESSI optics,
which applies in situations where imaging information is less in demand than (spatially integrated)
temporal evolution. The prior assumptions involve time-independence of the true brightness 
distribution during $\tau$ $\ll$ $T_S$, and an a gaussian a priori pdf for the source density on the 
solar disc. The estimator minimizes the expected quadratic deviation of true and retrieved unmodulated 
counts, while enforcing agreement of their expectation values. Non-overlapping time intervals $\tau$ are
independent. Geometrically, the algorithm tries to cancel
the spatial transmission patterns (modulation patterns) of the RHESSI optics by a suitable linear 
combination of patterns belonging to the time interval $\tau$. The degree to which canceling is
beneficial depends on the counting noise, and the algorithm constructs a trade-off between Poisson-
and modulational uncertainties of the estimator. Monte Carlo simulations show that 
the mean relative error of the demodulation reaches the Poisson limit, and demonstrate
robustness against violation of the prior assumptions. An application 
to a solar eruption is also discussed.

The present method is limited in several ways. First, any non-solar background
is neglected. Secondly, the use of sharp time intervals $\tau$ brings along the computational 
advantage that the data can be split and the results merged in the end, but at the cost of
possible artifacts at interval boundaries. The use of larger and smoothly tapered time intervals
would help, but is computationally demanding.

\section{Appendix: Condition number inequalities}


The condition number of a positive definite matrix is defined as the ratio of its largest
eigenvalue to its smallest eigenvalue. We first ask for simple bounds on the condition number of the 
matrix ${\sf \Lambda}_{\mu \nu} = \lambda_\mu \lambda_\nu +\lambda_\mu \delta_{\mu \nu}$ with
positive $\lambda_\mu$. Since ${\sf \Lambda}$ is symmetric all its eigenvalues
ly between the minimum and maximum of the Rayleigh quotient $R = ({\bf \hat{x}}, {\sf \Lambda} {\bf \hat{x}})
= ({\bf \hat{x}} \cdot \blambda)^2 + \sum_\mu \lambda_\mu \hat{x}_\mu^2$ with $|{\bf \hat{x}}|^2 = 1$
(Euclidian norm). The minumum of $R$ is bounded by $R \ge \min (\sum_\mu \lambda_\mu \hat{x}_\mu^2) = \min (\lambda_\mu)$,
while the maximum is bounded by $R \le |{\bf \hat{x}}|^2 |\blambda|^2 + \max (\sum_\mu \lambda_\mu \hat{x}_\mu^2)
= |\blambda|^2 + \max(\lambda_\mu)$. Therefore, $\cond ({\sf \Lambda}) \le \big(|\blambda|^2+\max(\lambda_\mu)\big)
/\min(\lambda_\mu)$. 

Next we consider the matrix $\EB {\sf \Lambda}$. Since it represents an average
over different ${\sf \Lambda}$ with different principal directions,
we may expect that 

\begin{equation}
\cond(\EB{\sf \Lambda}) \le \max_B \cond({\sf \Lambda}) \, .
\label{cond_EB}
\end{equation}

To show that this is true it suffices to show that for any two positive definite matrices $A_1$ and $A_2$ of equal size
the inequality 

\begin{equation}
\cond(A_1+A_2) \le \max \big(\cond (A_1), \, \cond (A_2) \big)
\label{ineq}
\end{equation}

holds. Equation (\ref{cond_EB}) then follows by induction over $i$ with $A_i = dP(B_i) {\sf \Lambda}(B_i)$.
(Assuming that the true scenes may be labeled by discrete labels $i$.
We shall not prove this assertion, but call it plausible in view of the finite resolution of the
modulation patterns.) Equation (\ref{ineq}) is then easily verified by direct calculation:

\begin{eqnarray*}
\cond (A_1+A_2) & = & \frac{ \max ((\xhat, A_1 \xhat) + (\xhat, A_2 \xhat))}{\min ((\xhat, A_1 \xhat) + (\xhat, A_2 \xhat))} \\
	& \le & \frac{ \max (\xhat, A_1 \xhat) + \max (\xhat, A_2 \xhat)}{\min (\xhat, A_1 \xhat) + \min (\xhat, A_2 \xhat)} \\
	& = & \frac{\cond (A_1) \min (\xhat, A_1 \xhat) + \cond (A_2) \min (\xhat, A_2 \xhat)}{\min (\xhat, A_1 \xhat) + \min (\xhat, A_2 \xhat)} \\
	& \le & \max ( \cond(A_1), \, \cond(A_2) ) \, .
\end{eqnarray*}

The last line follows by either assuming $\cond (A_1)$$<$$\cond (A_2)$ or $\cond (A_1)$$>$$\cond (A_2)$.

\bigskip

The author thanks G. Hurford, A. Csillaghy, R. Schwartz, A. Benz, C. Wigger, P. StHilaire
and E. Kirk for helpful discussions and comments.

\bigskip

\begin{harvard}
\item[]Aschwanden M J, Benz A O and Schwartz R 1993 {\it ApJ}, {\bf 417} 790 
\item[] Byrne C Haughton D and Jiang T 1993 {\it Inverse Problems} {\bf 9} 39
\item[] Dennis B 1985 {\it Solar Phys.} {\bf 100} 465
\item[] Evans S N and Stark P B 2002 {\it Inverse Problems} {\bf 18} R55
\item[] Fivian M and Zehnder A 2002 {\it Solar Phys.} {\bf 210} 87
\item[] Golub G H and Van Loan C F 1989 {\em Matrix Computations} John Hopkins Univ. Press, Baltimore
\item[] Hurford G G et al. 2002a {\it Solar Phys.} {\bf 210} 61 
\item[] Hurford G G and D W Curtis 2002b {\it Solar Phys.} {\bf 210} 101
\item[] Lin R et al 2002 {\it Solar Phys.} {\bf 210} 2
\item[] Machado M E, Ong K, Emslie A G, Fishman G J, Meegan C, Wilson R and Paciesas W S 1993 
	{\it Adv. Space Res.} {\bf 123} 183
\item[] Press W H, Teukolsky S A Vettering W T and Flannery B P 1998 Numerical recipes in C,
2nd edition, Cambridge Univ. Press
\item[] Prince T A, Hurford G J Hudson H S and Cranell C J 1988 {\it Solar Phys.} {\bf 118} 269
\item[] Rybicki G B and Press W H 1992 {\it Astrophys. J.} {\bf 398} 169
\item[] Schnopper H W, Thompson R I, and S Watt 1968 {\it Space Sci. Rev.} {\bf 8} 534                                                                                              
\item[] Skinner G K 1997 {\it J. British Interplanetary Soc.} {\bf 32} 77
\item[] Skinner G K and T J Ponman 1995 {\it Inverse Problems} {\bf 10} 655
\item[] Smith D et al. 2002 {\it Solar Phys.} {\bf 210} 33
\item[] Willmore A P 1970 {\it Mon. Not. R. astr. Soc.} {\bf 147} 387
\item[] Zehnder et al 2003 Proceedings of the SPIE {\bf 4853} 41

\end{harvard}


\clearpage

\begin{figure}
\centerline{\epsfig{file=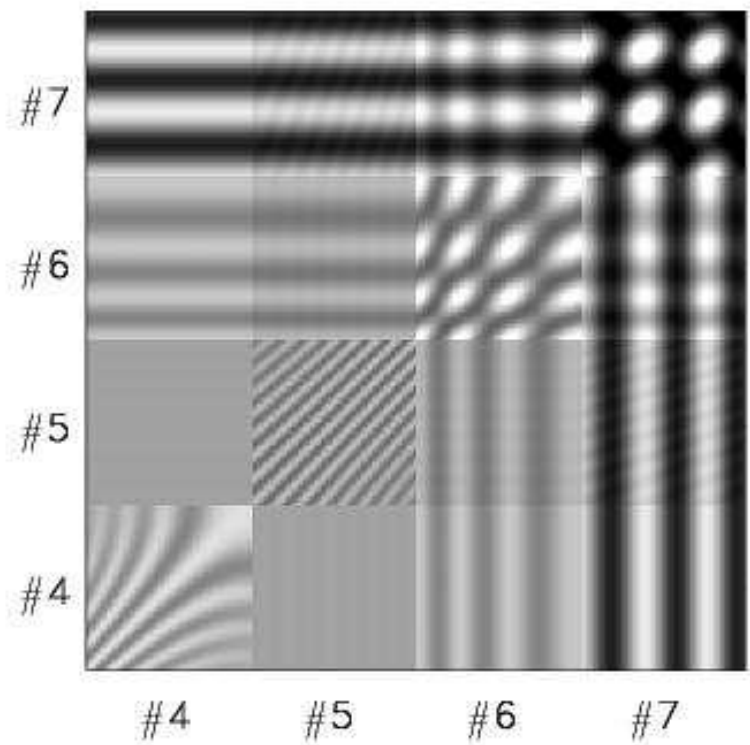,height=13cm} }
\caption{Correlation $\langle M_\mu M_\nu \rangle$ for 2.75s $\le$ $t$ $\le$ 3s in
Fig. \protect\ref{cw_fig}. Diagonal blocks contain the
temporal autocorrelations of the 4 subcollimators, with time running from left to right.
}
\label{mm_fig}
\end{figure}

\end{document}